\documentclass[12pt]{iopart}

\usepackage{graphicx}
\usepackage{amssymb}

\newcommand{ \ybco }{\mbox{YBCO$_{6+x}$}}
\newcommand{ \lsco }{\mbox{LSCO}}
\newcommand{ \ybcosixthree }{\mbox{YBCO$_{6.3}$}}
\newcommand{ \ybcosixthreefive }{\mbox{YBCO$_{6.35}$}}

\newcommand{ \ybcosixfourfive }{\mbox{YBCO$_{6.45}$}}
\newcommand{ \ybcosixfive }{\mbox{YBCO$_{6.5}$}}
\newcommand{ \ybcosixsix }{\mbox{YBCO$_{6.6}$}}
\newcommand{ \cuotwo }{\mbox{CuO$_2$}}

\def\vwu#1#2{\mbox{{#1}\,{#2}}}

\begin{document}

\title[Neutron scattering study of the magnetic phase diagram of underdoped YBa$_2$Cu$_3$O$_{6+x}$]{Neutron scattering study of the magnetic phase diagram of underdoped YBa$_2$Cu$_3$O$_{6+x}$}

\author{D Haug$^1$, V Hinkov$^1$, Y Sidis$^2$, P Bourges$^2$, N B Christensen$^{3,4,5}$, A Ivanov$^6$, T Keller$^{1,7}$, C T Lin$^1$ and B Keimer$^1$}

\address{$^1$ Max-Planck-Institut f{\"u}r Festk{\"o}rperforschung, Heisenbergstra{\ss}e 1, D-70569 Stuttgart, Germany}
\address{$^2$ Laboratoire L{\'e}on Brillouin, CEA-CNRS, CE-Saclay, F-91191 Gif-s{\^u}r-Yvette, France}
\address{$^3$ Laboratory for Neutron Scattering, ETH Z{\"u}rich \& Paul Scherrer Institut, CH-5232 Villigen, Switzerland}
\address{$^4$ Materials Research Division, Ris{\o} DTU, Technical University of Denmark, DK-4000 Roskilde, Denmark}
\address{$^5$ Nano-Science Center, Niels Bohr Institute, University of Copenhagen, DK-2100 Copenhagen, Denmark}
\address{$^6$ Institut Laue-Langevin, 6 Rue Jules Horowitz, F-38042 Grenoble Cedex 9, France}
\address{$^7$ ZWE FRM-II, Technische Universit{\"a}t M{\"u}nchen, Lichtenbergstra{\ss}e 1, D-85748 Garching, Germany}

\ead{\mailto{B.Keimer@fkf.mpg.de}}

\begin{abstract}
We present a neutron triple-axis and resonant spin-echo spectroscopy study of the spin correlations in untwinned
YBa$_2$Cu$_3$O$_{6+x}$ single crystals with $x= 0.3$, 0.35, and 0.45 as a function of temperature and magnetic field.
As the temperature $T \rightarrow 0$, all samples exhibit static incommensurate magnetic order with propagation vector along the
$a$-direction in the \cuotwo\ planes. The incommensurability $\delta$ increases monotonically with hole concentration,
as it does in La$_{2-x}$Sr$_x$CuO$_4$. However, $\delta$ is generally smaller than in La$_{2-x}$Sr$_x$CuO$_4$ at the
same doping level, and there is no sign of a reorientation of the magnetic propagation vector at the lowest doping levels.
The intensity of the incommensurate Bragg reflections increases linearly with magnetic field for
YBa$_2$Cu$_3$O$_{6.45}$ (superconducting $T_\mathrm{c} = 35$ K), whereas it is field-independent for
YBa$_2$Cu$_3$O$_{6.35}$ ($T_\mathrm{c} = 10$ K). These results fit well into a picture in which superconducting and
spin-density-wave order parameters coexist, and their ratio is controlled by the magnetic field. They also suggest that
YBa$_2$Cu$_3$O$_{6+x}$ samples with $x \sim 0.5$ exhibit incommensurate magnetic order in the high fields used for the recent quantum
oscillation experiments on this system, which likely induces a reconstruction of the Fermi surface.
We present neutron resonant spin-echo measurements (with energy resolution $\sim 1$
$\mu$eV) for $T \neq 0$ that demonstrate a continuous thermal broadening of the incommensurate magnetic Bragg reflections into a quasielastic
peak centered at excitation energy $E=0$, consistent with the zero-temperature transition expected for a two-dimensional
spin system with full spin-rotation symmetry. Measurements on YBa$_2$Cu$_3$O$_{6.45}$
with a conventional triple-axis spectrometer (with energy resolution $\sim 100$
$\mu$eV) yield a characteristic crossover temperature $T_{\rm SDW} \sim 30$ K for the onset of quasi-static magnetic
order. Upon further heating, the wavevector characterizing low-energy spin excitations progressively approaches the
commensurate antiferromagnetic wave vector, and the incommensurability vanishes in an order-parameter-like fashion at
an ``electronic liquid-crystal'' onset temperature $T_{\rm ELC} \sim 150$ K. Both $T_{\rm SDW}$ and $T_{\rm ELC}$
increase continuously as the Mott-insulating phase is approached with decreasing doping level. These findings are
discussed in the context of current models of the interplay between magnetism and superconductivity in the cuprates.
\end{abstract}

\maketitle

\section{Introduction}

The phase diagram of the cuprate high-temperature superconductors includes three generic thermodynamic phases at zero
temperature: a Mott-insulating phase with commensurate antiferromagnetic order, a $d$-wave superconducting phase, and a
metallic phase at low, intermediate, and high doping levels, respectively. In the limits of zero and large doping
levels, the low-energy excitations of these phases are well understood. In the Mott-insulator, spin and charge
excitations are separated by a large optical gap, and neutron scattering experiments have shown that the low-energy
spin excitations are well described by a Heisenberg spin Hamiltonian \cite{Tranquada89,Coldea01}. In the metallic phase
at high doping levels, various transport experiments indicate that the Fermi liquid theory yields an adequate
description of the coupling between spin and charge excitations \cite{Proust02,Nakamae03,Vignolle08}. In the $d$-wave
superconducting (SC) phase, however, a quantitative description of the coupling between spin and charge excitations
remains a challenge for current research, and there are some indications that such a description will also answer the
central question about the origin of high-temperature superconductivity.

The recent discovery of quantum oscillations in YBa$_2$Cu$_3$O$_{6+x}$ (\ybco) with $x \sim 0.5$ and in $\rm YBa_2 Cu_4
O_8$ indicates that the Fermi liquid theory may also be applicable to underdoped cuprates
\cite{Doiron07,LeBoeuf07,Yelland08,Bangura08,Sebastian08}, but the small size of the observed pockets is inconsistent
with the large hole-like Fermi surface predicted by ab-initio electronic structure calculations. This suggests a
reconstruction of the Fermi surface by an electronic superstructure, at least in the presence of the high magnetic
fields required for the quantum oscillation experiments. An incommensurate magnetic superstructure has indeed been
observed in underdoped La$_{2-x}$Sr$_x$CuO$_4$ (\lsco) \cite{Yamada98,Fujita02,Wakimoto99,Wakimoto00}, but until
recently had appeared to be a peculiar property of this family of compounds. Moreover, quantum oscillations have not yet been observed in \lsco, presumably due to a higher level of intrinsic disorder.

We recently reported a related
superstructure in a \ybco\ crystal with $x = 0.45$ \cite{Hinkov08}, slightly below the doping level at which the
quantum oscillations were observed. Although the incommensurability is about a factor of two smaller than the one in
\lsco\ at the same doping level, this implies that incommensurate magnetic order should be regarded as another generic
zero-temperature phase in the cuprate phase diagram. It remains to be established whether the incommensurate Bragg
reflections observed by neutron scattering are due to a modulation of the amplitude (``stripes'') \cite{Kivelson03} or
the direction (``spirals'') \cite{Sushkov05} of the ordered moments, or both. Nonetheless, we refer to this state as a
``spin density wave'' (SDW), taking account of the observation that the incommensurate magnetic order in
\ybcosixfourfive\ develops out of a metallic state, in contrast to the commensurate antiferromagnetic order in the
Mott-insulating state at lower doping levels. The amplitude of the SDW modulation is strongly enhanced by a modest
magnetic field $\textbf{H} \sim 15$ T \cite{Haug09}. It is thus very likely that the same SDW state is present at the
much higher fields used in the quantum oscillation experiments on crystals with slightly higher doping levels. A
reconstruction of the Fermi surface by the SDW is then expected on general grounds. Subsequent work has confirmed that
the size of the Fermi surface pockets observed in \ybcosixfive\ can be quantitatively explained based on the
incommensurate wave vector we have determined \cite{Harrison09,Sebastian10}.

An investigation of the temperature dependence of the magnetic structure and dynamics of \ybcosixfourfive\ has revealed
that the magnetic order disappears in a two-step fashion upon heating \cite{Hinkov08}. In a first step, muon spin
relaxation ($\mu$SR) experiments showed that static magnetic order, which reduces the translational symmetry of the
lattice, vanishes at a temperature $T \sim 2$ K. In a second step, the incommensurate wave vector characterizing
low-energy magnetic fluctuations continuously approaches the commensurate antiferromagnetic ordering wave vector with
increasing temperature, and the incommensurability vanishes in an order-parameter-like fashion at $T_{\rm ELC} \sim
150$ K. While the magnetic fluctuations above $T_{\rm ELC}$ respect the fourfold rotation symmetry of the CuO$_2$
plane, their wave vector spontaneously aligns with one of the two nearly equivalent copper-oxygen bond directions upon
cooling below this temperature. In a wide temperature regime, therefore, the spin system collectively behaves in such a
way that the rotational symmetry is broken while the translational symmetry remains intact. Since the symmetry
properties of this state are analogous to those of a nematic liquid crystal, we refer to this state as an ``electronic
liquid-crystal'' (ELC) \cite{Kivelson98}. The small orthorhombicity of the \ybcosixfourfive\ crystal lattice is required to align the ELC domains such
that the ELC state becomes observable in a volume-integrating method such as neutron scattering. In the presence of
this aligning field, the ELC transition is rounded into a sharp crossover. While the neutron scattering data are not
accurate enough to quantify this rounding, the strong, order-parameter-like temperature dependence of the
incommensurability indicates that the loss of rotational symmetry reflects an underlying phase transition that is
driven by collective interactions between the spins.

Our work on \ybcosixfourfive\ raises several questions. How is the incommensurate SDW related to the magnetic order
observed in \lsco, and how does it evolve into the commensurate antiferromagnetic order \cite{Tranquada89,Coneri10}
observed in \ybco\ at lower doping levels? How is the ELC state related to anomalies in the charge dynamics observed by
transport (such as Nernst effect measurements \cite{Wang06,Daou10}) and spectroscopic probes (such as infrared spectroscopy
\cite{Timusk99}) in the same temperature and doping range? This article describes neutron scattering measurements on
untwinned \ybco\ crystals with $x=0.3$ and $0.35$ designed to address these questions, and to constrain theoretical
models of the collective magnetic ordering phenomena in the cuprates. In conjunction with measurements on $x=0.45$ and
higher doping levels, the results allow us to determine the doping dependence of the incommensurate wave vector as well
as the characteristic crossover temperatures for SDW and ELC ordering in \ybco. For a \ybcosixthreefive\ crystal, we also present data on the
magnetic field dependence of the SDW order parameter at low temperatures, as well as temperature dependent neutron resonant spin-echo experiments that bridge
the gap in energy scales between $\mu$SR and conventional neutron diffraction data. The results of our study allow us to
draw an outline of the phase diagram of underdoped \ybco\ as a function of doping, magnetic field, and temperature.

\section{Experimental details}

The measurements were performed on crystal arrays with three different hole doping levels in the
strongly underdoped regime of the \ybco\ phase diagram. The samples are listed in Table \ref{tab:sampletable}.
The oxygen content of the single crystals was carefully adjusted by a thermal treatment during which oxygen
diffusion takes place. A variation of temperature at fixed oxygen partial pressure results in different oxygen contents
and thus different hole concentrations \cite{Jorgensen90}. The oxygen content and hole concentration per
planar Cu ion, $p$, were extracted from the known doping dependence of the out-of-plane lattice parameter $c$
and $T_\mathrm{c}$ \cite{Liang06}.

\begin{table}[h]
\begin {center}
\begin{tabular}{|c|c|c|c|c|}
\hline
Sample & $T_\mathrm{c}$ & lattice parameter $c$ ($\mathrm{\,\AA}$) & hole doping $p$ &  total mass $m_\mathrm{tot}$ (g) \\
\hline\hline
\ybcosixfourfive & \vwu{35}{K} & 11.761 & 0.082 & 2.0 \\
\ybcosixthreefive & \vwu{10}{K} & 11.781 & 0.062 & 1.4 \\
\ybcosixthree & \vwu{0}{K} & 11.791 & 0.052 & 1.0 \\
\hline
\end{tabular}
\vspace{0.5cm} \caption{Summary of the underdoped \ybco\ samples. The SC transition temperature
$T_\mathrm{c}$ was determined by magnetometry. The out-of-plane lattice parameter $c$ at room temperature was determined by X-ray powder
diffraction on selected samples from the same batch. From this value the hole doping level per planar Cu ion
$p$ could be extracted \cite{Liang06}. Several tens of crystals were co-aligned by X-ray Laue diffractometry,
mimicking a
large single-domain crystal with a total mass $m_\mathrm{tot}$.} \label{tab:sampletable}
\end{center}
\end{table}

In \ybco, a unique axis in the CuO$_2$ plane is established by the orthorhombic distortion of the crystal structure.
However, as-grown samples show crystallographic twinning, because the difference of the in-plane lattice parameters $a$
and $b$ is small. In order to observe the anisotropy of the electron system with a volume-averaging probe such as
neutron scattering, a single-domain state has to be prepared in which the orientation of the two in-plane axes is
maintained throughout the entire sample volume. This was accomplished by individually detwinning single crystals of
\ybco\ (with typical size $a \times b \times c = 2 \times 2 \times \vwu{0.5}{mm$^3$}$) by application of uniaxial
mechanical stress along the crystallographic $(1,0,0)$ direction.

The crystals were characterized by magnetometry and were found to exhibit SC transitions with widths of 2--\vwu{4}{K},
testifying to their high quality. In order to obtain a sample volume sufficient for inelastic neutron scattering
experiments, several tens of crystals were co-aligned on a single-crystalline silicon disc by X-ray Laue
diffractometry. The resulting crystal arrays, mimicking large single-domain crystals, had mosaicities of $\leq
1.5^{\circ}$ and a majority twin domain population close to 90\%. The contribution of the minority twin domains to the
observed magnetic scattering signal is thus below the detection limit.

The neutron experiments were performed at the triple-axis spectrometers IN8 and IN14 (ILL, Grenoble, France), 4F1 and
4F2 (LLB, Saclay, France), Rita-II (SINQ, PSI Villigen, Switzerland), and on the resonant spin-echo triple-axis
spectrometer TRISP (FRM-II, Garching, Germany). Pyrolytic-graphite crystals were used to monochromate and analyze the
neutron beam. In order to extinguish higher-order contaminations of the neutron beam, a beryllium filter was inserted
into the beam for those measurements which were performed with fixed final wave vector $k_\mathrm{f}=1.55
\mathrm{\,\AA}^{-1}$. For the same purpose a pyrolytic-graphite filter was used when measuring at $k_\mathrm{f}=2.57
\mathrm{\,\AA}^{-1}$ and $k_\mathrm{f}=2.66 \mathrm{\,\AA}^{-1}$, respectively. On TRISP, two radio-frequency coil
sets, located between the monochromator and sample and between sample and analyzer, were used to manipulate the Larmor
phase of the neutron spin before and after the scattering event, and the relaxation rate of magnetic excitations was
extracted from spin-echo profiles \cite{Bayrakci06}.

The scattering vector $\textbf{Q}=(H,K,L)$ is expressed in reciprocal lattice units (r.l.u.), i.e. in units of the
reciprocal lattice vectors $a^*$, $b^*$, and $c^*$ ($a^*=2\pi/a$, etc.). As in the undoped ($x = 0$) parent compound,
magnetic intensity is concentrated around the antiferromagnetic wave vector $\textbf{Q}_\mathrm{AFM}=(0.5,0.5, L)$.
Scans were thus performed either around $\textbf{Q}_\mathrm{AFM}$ or around the equivalent points $(0.5, 1.5, L)$ and
$(1.5, 0.5, L)$ in higher Brillouin zones.

For the measurements in high magnetic field the sample was mounted in a 15-T vertical-field cryomagnet. The scattering
plane was spanned by the vectors $(1,0,0)$ and $(0,1,2)$, and scans were performed along $a^*$ around $\textbf{Q}=(0.5,
0.5, 1)$. The scattering geometry implies an angle of 33$^{\circ}$ between the magnetic field $\textbf{H}$ and
the $c$-axis. Thus, for the field-dependent experiments, the external magnetic field has a major component
perpendicular to the \cuotwo\ planes. All scans for $\textbf{H} \neq 0$ were performed after field cooling.

\section{Results}

\subsection{Doping dependence}

\begin{figure}[t]
\begin{center}
\includegraphics[width=0.45\columnwidth,angle=0]{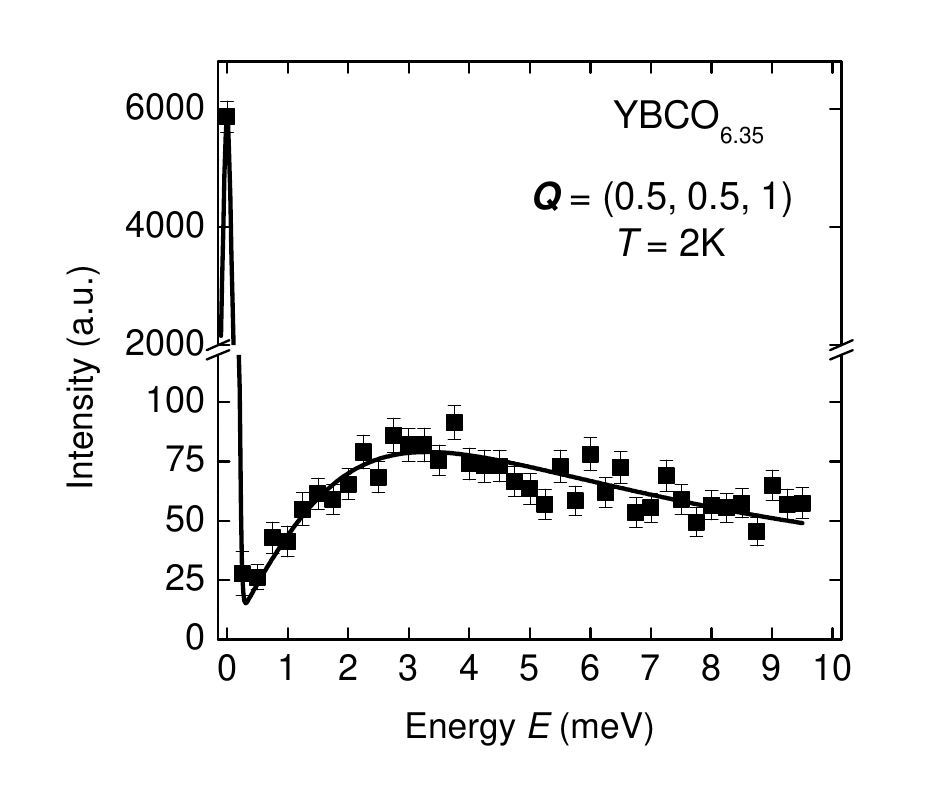}
\caption {Magnetic neutron scattering function of \ybcosixthreefive\ at $T=2$ K and $\textbf{Q}$ = (0.5, 0.5, 1), showing distinct quasielastic and inelastic components.} \label{figure1}
\end{center}
\end{figure}

We first discuss the behavior of the spin system at the lowest temperatures studied in our experiments ($T = 2$
K). Figure \ref{figure1} shows the low-temperature neutron scattering function $S(\textbf{Q},\omega$) of
\ybcosixthreefive\ around the antiferromagnetic ordering wave vector $\textbf{Q}_\mathrm{AFM}=(0.5,0.5,1)$ as a
function of excitation energy $E =\hbar \omega$. It is apparent that the spectrum is comprised of two components: a
sharp, intense ``quasi-elastic'' peak centered at $E =0$, and a weaker inelastic component that depends smoothly on
energy. The two components are analogous to the Bragg peak and the spin wave spectrum of an antiferromagnetically
ordered state. At the lowest temperatures, the energy width of the quasielastic peak is at the resolution limit of neutron resonant spin-echo spectroscopy ($\sim 1 \mu$eV, see below), and $\mu$SR experiments indicate magnetic order that persists much longer than the muon lifetime ($\sim 2 \mu$sec). At temperatures of order 2 K, which includes the
temperature range in which the quantum oscillation experiments were carried out, the quasielastic peak can therefore be
regarded as a signature of static magnetic order. The two-component form of the magnetic spectrum is common to all
three samples studied here\footnote{For \ybcosixthree\ and \ybcosixthreefive, both commensurate quasielastic
reflections centered at integer $L$ and incommensurate peaks with continuous $L$-dependence characteristic of
two-dimensional magnetic order are detected. Both sets of reflections exhibit a similar temperature dependence.}.

\begin{figure}[t]
\begin{center}
\includegraphics[width=0.9\columnwidth,angle=0]{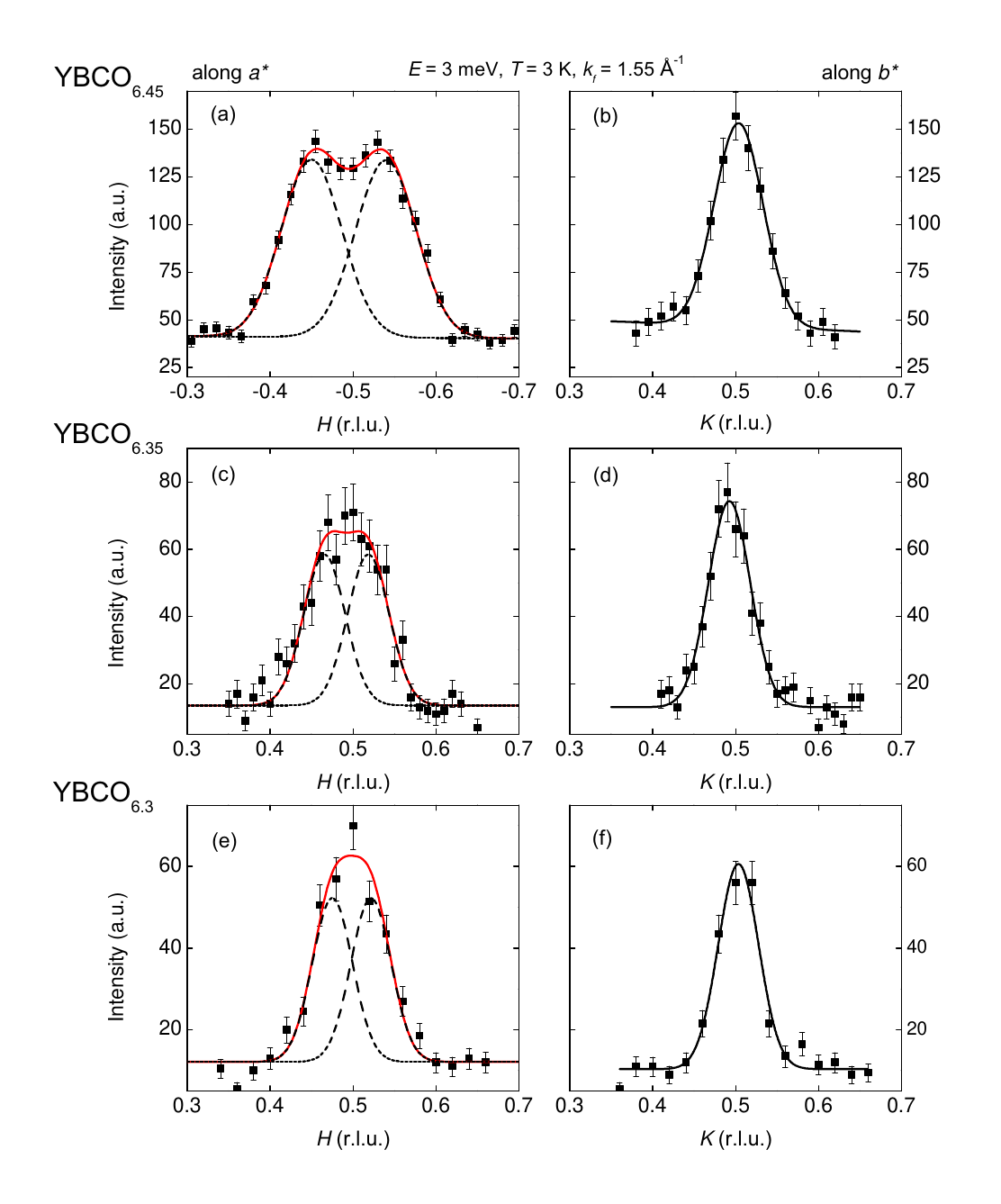}
\caption {Constant-energy scans (raw data, without background correction) at $E=$ \vwu{3}{meV} and $T=$ \vwu{3}{K}
along $a^*$ and $b^*$ for (a,b) \ybcosixfourfive, (c,d) \ybcosixthreefive,  and (e,f) \ybcosixthree. The lines are
single Gaussian ($b^*$) and double-Gaussian ($a^*$) profiles fitted to the data. The two incommensurate peaks
symmetrically displaced from $\textbf{Q}_\mathrm{AFM}$ along $a^*$ are shown as dashed lines.} \label{figure2}
\end{center}
\end{figure}

In order to determine the spatial character of the magnetic correlations, we have performed constant-energy scans through the
low-energy spin excitation spectrum. Figure \ref{figure2} shows a comparison of constant-energy scans for $E=$
\vwu{3}{meV} along the in-plane crystallographic directions $a^*$ and $b^*$ for \ybcosixfourfive, \ybcosixthreefive,
and \ybcosixthree. Note that these are raw data, without any background subtraction. Scans through the quasielastic
peak are of lower quality because they are superposed by a background due to incoherent scattering from the sample and
the sample mount, but the scan profiles are nearly identical \cite{Hinkov08}. Clearly, the magnetic response of all three
samples shows a pronounced in-plane anisotropy. The cuts through the \ybcosixfourfive\ spectrum (Fig. \ref{figure2}a,b)
demonstrate that the anisotropic intensity distribution is a consequence of two incommensurate peaks that are
symmetrically displaced from $\textbf{Q}_\mathrm{AFM}$ along $a^*$, whereas the distribution is commensurate along
$b^*$. For \ybcosixthreefive\ the anisotropy is qualitatively similar, but less pronounced (Fig. \ref{figure2}c,d), and
it is further reduced for \ybcosixthree\ (Fig. \ref{figure2}e,f). Note that even in this latter case we can rule out
any influence of the instrumental resolution on the anisotropic peak profile\footnote{The full-width-at-half-maximum of the
elastic resolution function was determined at $\textbf{Q}_\mathrm{AFM}=(0.5,0.5,1)$ using the $\lambda/2$ harmonic of
the Bragg reflection (1 1 2) by removing the filters; it is $\sim (0.025 \pm 0.002)$ r.l.u. in the $a^*$- and $\sim
(0.023 \pm 0.002)$ r.l.u. in the $b^*$-direction. The effective inelastic resolution is somewhat larger, but still much
smaller than the width of the scan profiles both along $a^*$ and $b^*$.}. Although the splitting between the two
incommensurate peaks in the $a^*$ direction is no longer resolved for \ybcosixthreefive\ and \ybcosixthree, fits to
two-Gaussian profiles with individual peak widths matching the one in the $b^*$ direction provide excellent descriptions of
the data (lines in Fig. \ref{figure2}). Together with the observation of a split profile for \ybcosixfourfive\ (Fig.
\ref{figure2}a), this indicates an incommensurability of the magnetic response that increases continuously with
increasing hole content in the \cuotwo\ planes.

\begin{figure}[t]
\begin{center}
\includegraphics[width=0.45\columnwidth,angle=0]{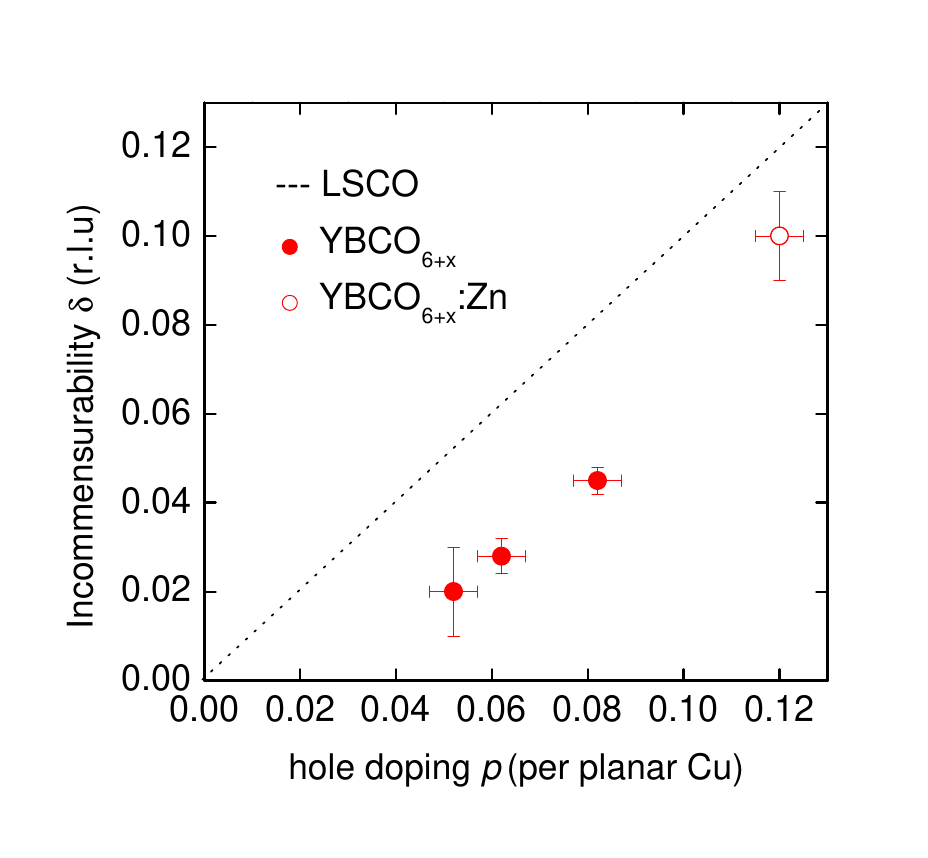}
\caption {Doping dependence of the incommensurability $\delta$ (i.e. the displacement of the peaks from
$\textbf{Q}_\mathrm{AFM}$) for an energy transfer of $E=$ \vwu{3}{meV} at low temperatures, extracted from the fits in
Fig. \ref{figure2} (closed symbols). The open symbol denotes the incommensurability of the low-energy spin excitations
for Zn-substituted \ybcosixsix\ (from \cite{Suchaneck10}). For comparison, the linear $\delta$-versus-$p$ relation for
\lsco\ is shown as a dashed line \cite{Yamada98,Fujita02}. } \label{figure3}
\end{center}
\end{figure}

Figure \ref{figure3} summarizes the doping dependence of the incommensurability $\delta$ in the three \ybco\ samples
that exhibit static magnetic order. We have also included data obtained from a \ybcosixsix\ sample in which static
incommensurate magnetic order was induced by substituting 2\% of the Cu atoms by spinless Zn impurities
\cite{Suchaneck10}. It is instructive to compare these data to the doping dependence of $\delta$ in the \lsco\ family
of cuprates, where the propagation vector of the incommensurate modulation of the spin system is along the Cu-O bond
direction for $p \geq 0.05$, as it is in \ybco\ \cite{Yamada98,Fujita02}. However, $\delta$ is systematically lower in
\ybco\ than it is in \lsco\ at the same $p$. While in \lsco\ a linear relationship between the incommensurability and
the doping level ($\delta = p$) is observed for $p \lesssim 0.12$, $\delta$ in \ybco\ extrapolates to zero as $p
\rightarrow 0.04$ from above, close to the doping level at which the propagation vector switches from the Cu-O bond
direction to the diagonal of a CuO$_4$ plaquette in \lsco\ \cite{Wakimoto99,Wakimoto00}. In contrast, we have not found
any evidence for a 45$^\circ$ reorientation of the incommensurate modulation in \ybco\ in the doping range we have
investigated. Possible origins of this materials-specific behavior include exchange interactions between directly
adjacent copper oxygen planes in the crystallographic unit cell (which is comprised of bilayer units in \ybco\ and a
single-layer unit in \lsco), and the lower disorder in \ybco\ which allows metallic conduction at lower $p$ than in
\lsco. Calculations considering the impact of these factors in the framework of a model in which the magnetic
incommensurability arises from spiral order are indeed consistent with the behavior we have observed \cite{Sushkov09}.

\subsection{Magnetic field dependence}

As discussed in Section 1, we have shown that the intensity of the quasielastic peak in \ybcosixfourfive\ is strongly
enhanced in an external magnetic field, while its shape is only weakly affected \cite{Haug09}. This effect is analogous
to a similar phenomenology previously established in \lsco\ for $x \sim 1/8$ \cite{Katano00,Lake02,Khaykovich05}. It is
naturally explained as a consequence of a competition between SDW and $d$-wave superconducting phases, which coexist in
\ybcosixfourfive\ for $\textbf{H}=0$ \cite{Demler01}. Since the magnetic field destabilizes $d$-wave
superconductivity by orbital depairing, the SC order parameter is reduced, and the SDW order parameter consequently
enhanced for $\textbf{H} \neq 0$.

\begin{figure}[t]
\begin{center}
\includegraphics[width=0.45\columnwidth,angle=0]{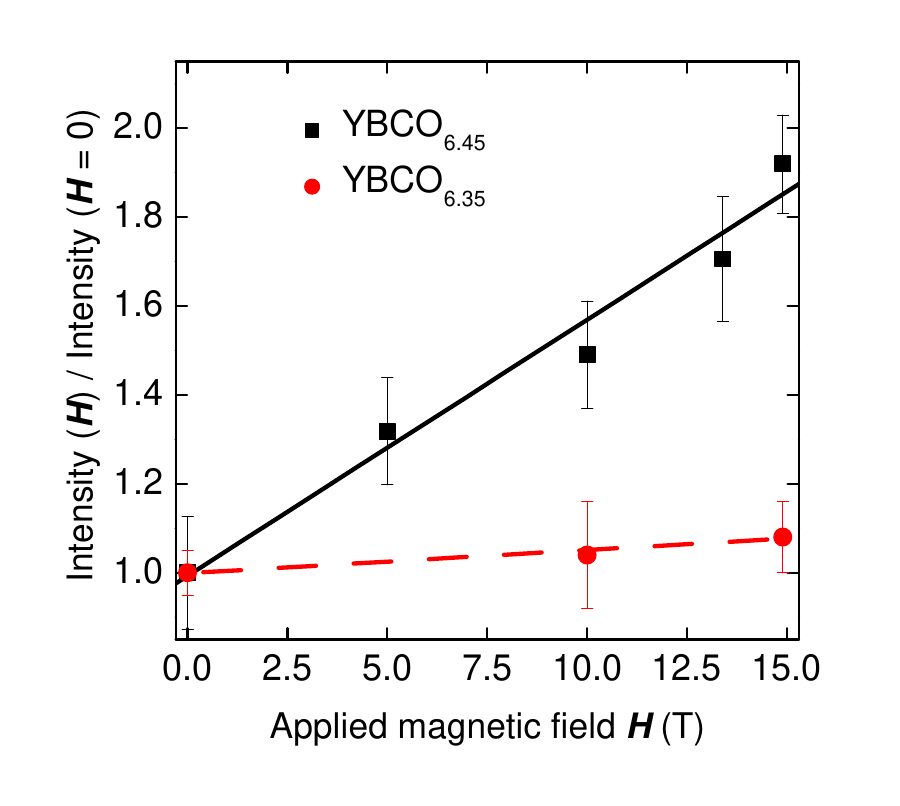}
\caption {Magnetic field dependence of the elastic magnetic peak at $T=$ \vwu{2}{K} for \ybcosixfourfive\ (reproduced
from \cite{Haug09}) and \ybcosixthreefive. The data were normalized to the intensity at $\textbf{H}=0$. }
\label{figure4}
\end{center}
\end{figure}

In order to further explore the validity of this scenario, we have extended our high-magnetic-field measurements to
\ybcosixthreefive. Fig. \ref{figure4} shows the $\textbf{H}$-dependence of the quasielastic peak intensity at
$T=$ \vwu{2}{K} for the two different doping levels. Whereas the peak intensity for \ybcosixfourfive\ increases linearly
with increasing magnetic field \cite{Haug09}, the magnetic intensity is field-independent within the error bar for
\ybcosixthreefive. The field-independence is consistent with earlier observations on a twinned \ybco\ sample with
similar hole concentration \cite{Stock09} and on \lsco\ samples in which the SDW order is already well established for
$\textbf{H}=0$ \cite{Chang08}. It also fits well into a two-phase competition scenario in which the SDW phase
fraction in \ybcosixthreefive\ is already close to 100\% for $\textbf{H}=0$.

\subsection{Temperature dependence}

\subsubsection{Spin density wave}

\begin{figure}[t]
\begin{center}
\includegraphics[width=0.95\columnwidth,angle=0]{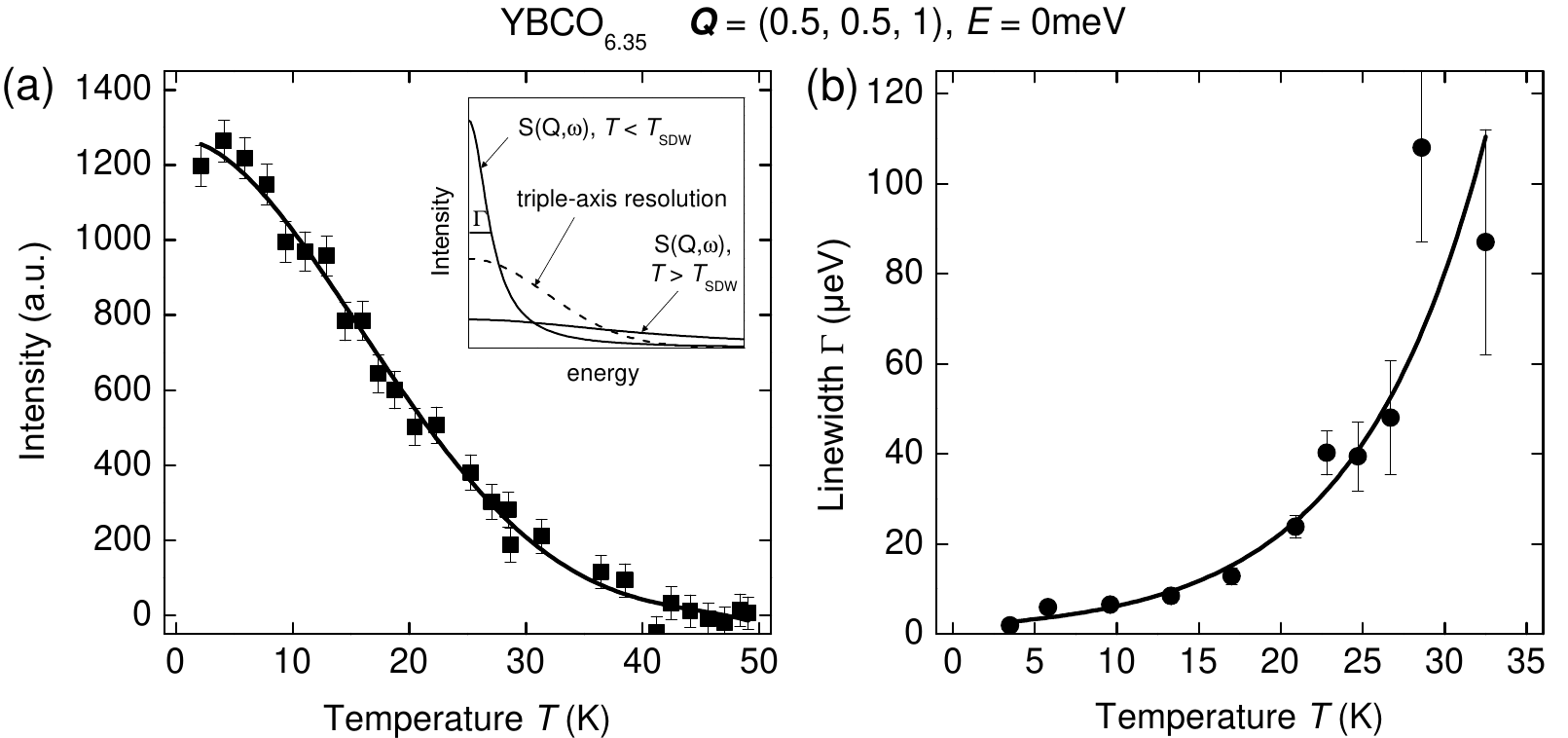}
\caption {Intensity of the quasielastic peak in \ybcosixthreefive\ determined by conventional cold-neutron triple-axis
spectrometry (a) and energy width (HWHM) of the same peak determined by resonant spin-echo spectroscopy (b) as a
function of temperature. The lines are guides-to-the-eye. The inset shows a sketch of the Lorentzian lineshape of the
quasielastic peak determined by spin-echo spectroscopy at different temperatures (solid lines) in relation to the
Gaussian resolution of the triple-axis spectrometer (dashed line).} \label{figure5}
\end{center}
\end{figure}

We now address the temperature dependence of the spin correlations, beginning with the intensity of the quasielastic
peak. Figure \ref{figure5}a shows the temperature dependence of the quasielastic peak intensity of \ybcosixthreefive\
measured on a conventional cold-neutron triple-axis spectrometer with energy resolution $\sim
0.1$ meV (half-width-at-half-maximum, HWHM). The data appear to show a SDW transition at a temperature of $T_\mathrm{SDW} \sim 40$ K. However, the
rounding of the transition, which substantially exceeds the width of the SC transition in these samples ($\sim$ 2--4
K), and the lack of saturation at low temperatures are unusual. Moreover, $\mu$SR experiments indicate ``spin
freezing'' transitions at much lower temperatures in this range of the phase diagram
\cite{Hinkov08,Coneri10,Niedermayer98}. Similar observations have been reported for underdoped \lsco\
\cite{Sternlieb90,Keimer92}.

In order to explore the origin of this behavior, we have carried out neutron resonant spin-echo measurements
\cite{Bayrakci06} with an energy resolution of $\sim 1$ $\mu$eV, which bridge the gap in energy scales between
conventional neutron diffraction and $\mu$SR. The spin-echo profiles indicate that the quasielastic peak exhibits a
Lorentzian lineshape with an intrinsic energy width (HWHM), $\Gamma$, that exceeds the resolution of the spin-echo
instrument for $T \geq 5$ K (Fig. \ref{figure5}b). $\Gamma$ increases continuously with increasing temperature and
exceeds the HWHM of the Gaussian resolution function of the triple-axis instrument for $T \geq 40$ K (inset in Fig.
\ref{figure5}a). These results show that the rounding observed in Fig. \ref{figure5}a is of dynamic origin, and not due
to an inhomogeneous distribution of transition temperatures as one might have suspected. They also imply that the SDW
transition suggested by Fig. \ref{figure5}a is actually a crossover whose specific form is determined by the energy
window sampled by the experimental method. For comparison with the results of transport experiments and other
experimental probes of the charge dynamics, it is nonetheless useful to monitor the strength of the quasi-static SDW
correlations at $T \neq 0$, which can be characterized by the crossover temperature $T_\mathrm{SDW}$ above which the
width of the quasielastic peak exceeds the energy scale $\sim 0.1$ meV. We have therefore extracted this temperature
from the data of Fig. \ref{figure5}a and analogous data collected for the other doping levels.

The diverging fluctuation time scale implied by the data of Fig. \ref{figure5}b is naturally explained as a consequence
of the zero-temperature phase transition expected for a two-dimensional spin system with full spin-rotation symmetry. In this scenario, the
intensity of the quasielastic peak for $T \rightarrow 0$ is proportional to the SDW order parameter. In contrast to the
divergence of the correlation length expected for a $T=0$ phase transition, the width of the scattering profiles in
momentum space (Fig. \ref{figure2}) saturates at low temperature. This may reflect either an inhomogeneous broadening
of the profiles due to slight variation of doping levels across the large single-crystal array, or finite-size effects
imposed by the random distribution of dopant atoms and/or the coexistence of SDW and SC phases. We note, however, that
the two-component form of the scattering function shown in Fig. \ref{figure1} is closely similar to the spectrum of an
undoped two-dimensional square-lattice antiferromagnet at $T \neq 0$, where thermal fluctuations generate a finite,
intrinsic correlation length \cite{Arovas88,Katanin10}. Possible origins for the intensity decrease of the inelastic component as
$\omega \rightarrow 0$ in the low-temperature data of Fig. \ref{figure1} include disorder-induced finite-size correlations 
and the intra-atomic spin-orbit coupling, which leads to anisotropy gaps in the spin-wave spectra of undoped cuprates \cite{Tranquada89}.

\subsubsection{Electronic liquid crystal}

At temperatures exceeding $T_{\rm SDW} \sim 40$ K, the quasielastic peak of Fig. \ref{figure1} is no longer visible,
but prominent low-energy incommensurate spin fluctuations persist. Since the incommensurability could not be resolved
in most of the inelastic scattering profiles for $x \leq 0.35$, we have extracted the momentum widths (FWHM) of the
profiles in the $a^*$ and $b^*$ directions, $\Delta_a$ and $\Delta_b$, from scans like those shown in Fig.
\ref{figure2} at different temperatures. Whenever the incommensurate peak separation $2 \delta$ exceeded $\Delta_b$,
$\Delta_a$ was extracted from a two-Gaussian fit. Otherwise the $a^*$ scan was fitted by a single Gaussian for the
sake of simplicity. For either way of fitting, $\Delta_a$ is the difference of the $H$ values for which the intensity
drops to half of the maximum intensity.

\begin{figure}[t]
\includegraphics[width=1\columnwidth,angle=0]{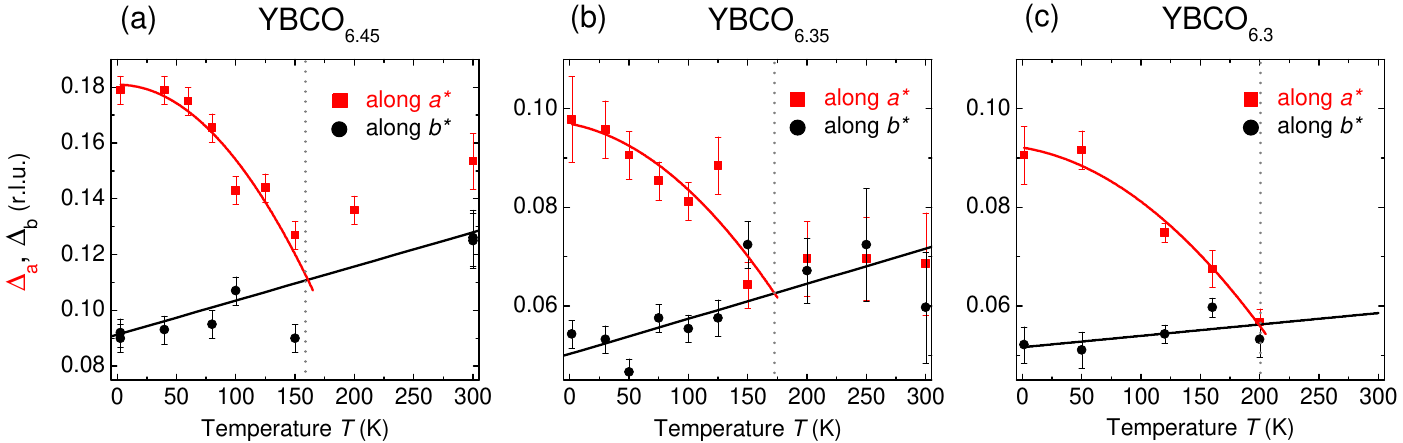}
\caption {Temperature evolution of the effective momentum widths (FWHM) along the $a$ and $b$ directions of the spin
excitation profiles at $E=$ \vwu{3}{meV} for (a) \ybcosixfourfive\ (b) \ybcosixthreefive, and (c) \ybcosixthree. The
widths were extracted from constant-energy scans, as shown in Fig. \ref{figure2}, and corrected for the instrumental
resolution.} \label{figure6}
\end{figure}

Figure \ref{figure6} shows the resulting widths $\Delta_a$ and $\Delta_b$, corrected for the instrumental resolution.
While $\Delta_b$ increases gradually upon heating, $\Delta_a$ decreases and continuously approaches $\Delta_b$. The
temperature dependence of $\Delta_a$ was therefore fitted by a polynomial function, while for $\Delta_b$ a simple
linear fit was applied which accounts for the thermal broadening of the peaks. The point of intersection of these two
curves marks $T_{\rm ELC}$, the onset temperature of the incommensurability. Above $T_\mathrm{ELC}$, both widths are
nearly identical and increase in a parallel manner upon further heating.

\begin{figure}[t]
\begin{center}
\includegraphics[width=0.45\columnwidth,angle=0]{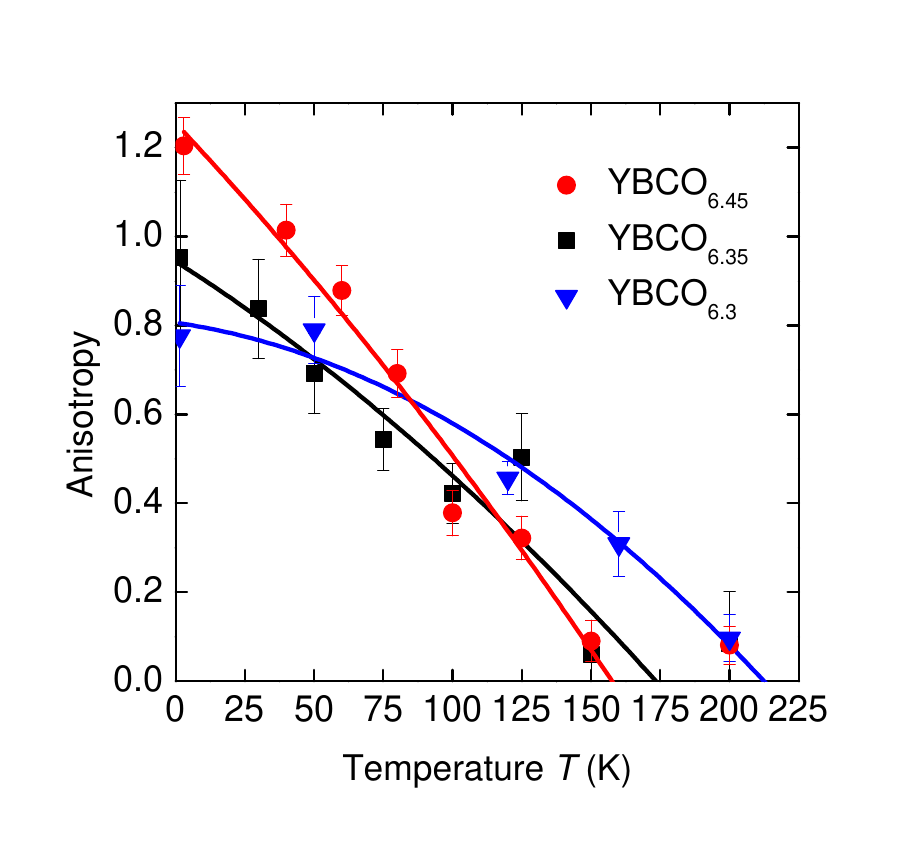}
\caption {Comparison of the temperature-dependent anisotropy, $\Delta_a / \Delta_b -1$, of the effective momentum widths $\Delta_a$ and $\Delta_b$ for
$E=$ \vwu{3}{meV} for the three different hole doping levels. The temperature for which the polynomial fit yields 
$\Delta_a / \Delta_b -1 =0$ is defined as the ELC transition temperature $T_\mathrm{ELC}$.} \label{figure7}
\end{center}
\end{figure}

In order to compare the three different samples directly, we have plotted the temperature dependence of the anisotropy,
defined as the deviation from unity of the ratio of $\Delta_a$ and the linear fit of $\Delta_b$ (Fig. \ref{figure7}). While the low-temperature
limit of the anisotropy decreases with decreasing hole concentration, reflecting the doping dependence of the
incommensurability (Fig. \ref{figure2}), the temperature where $\Delta_a / \Delta_b -1 \rightarrow 0$ increases with
decreasing $p$.

The strong temperature dependence of the in-plane anisotropy of the spin fluctuations in all three samples and its
appearance at a well-defined, doping dependent onset temperature confirms that this phenomenon arises from collective
interactions between spins. The ``Ising nematic'' transition expected on general grounds for a two-dimensional
incommensurate magnet \cite{Kivelson98,Vojta09,Kee03,Huh08,Sun10}, possibly associated with a ``Pomeranchuk''
instability of the Fermi surface \cite{Metzner00,Yamase00,Yamase09}, provides an interesting qualitative explanation of
our observations, but a quantitative assessment of this scenario will have to await further theoretical progress.
Meanwhile, it is interesting to compare our \ybco\ data to analogous observations on the
Ba$_2$Fe$_{2-x}$(Co,Ni)$_{x}$As$_2$ system, where a strong, doping dependent in-plane anisotropy of the spin
excitations has also been found \cite{Lester10,Inosov10}. In contrast to the cuprates, however, this anisotropy is
temperature independent and can be associated with nesting features of the Fermi surface \cite{Inosov10}. The SDW and
ELC states in \ybco, which are observed in immediate vicinity of the Mott insulator, thus do not appear to be
straightforwardly related to the more conventional SDW state in the iron pnictides.

\section{Summary}

\begin{figure}[t]
\begin{center}
\includegraphics[width=0.95\columnwidth,angle=0]{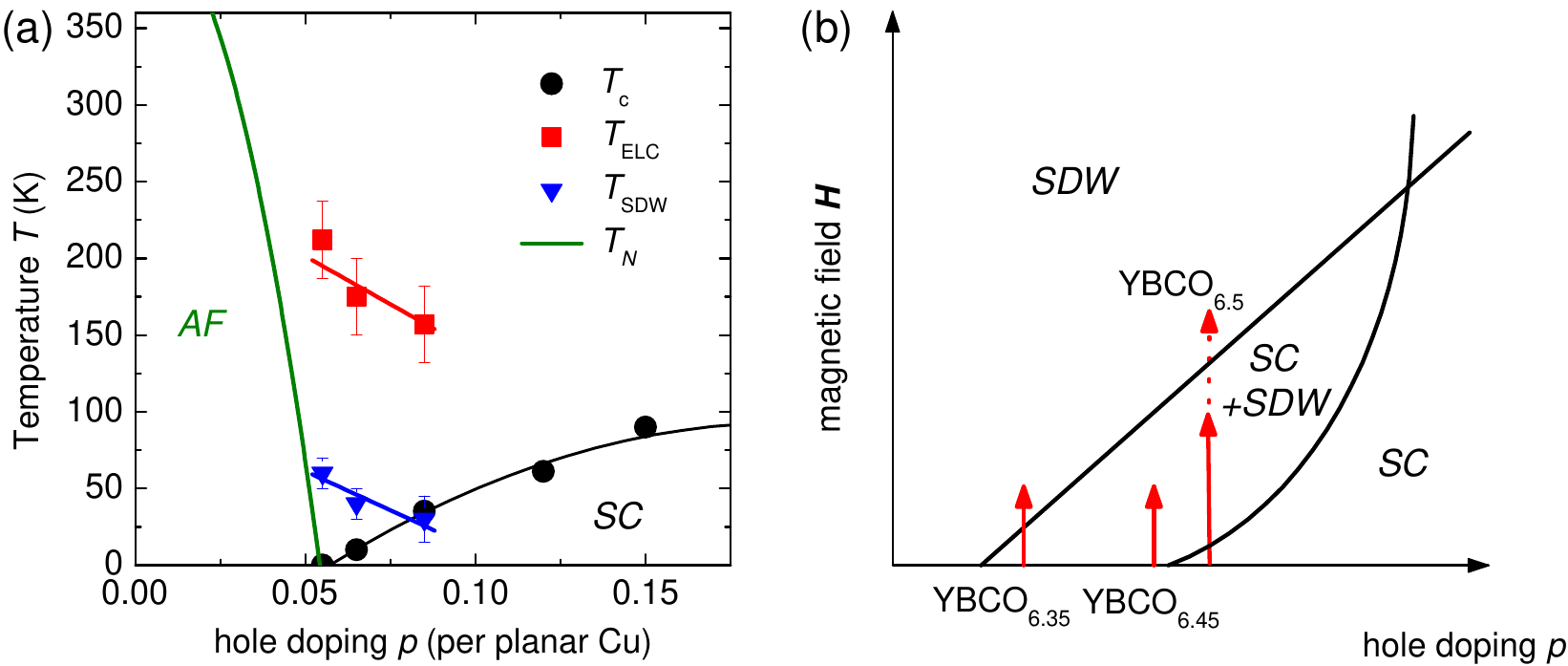}
\caption {Magnetic phase diagram of underdoped \ybco. Panel (a) shows a $p$--$T$ section of the phase diagram with the
crossover temperatures $T_\mathrm{SDW}$ and $T_\mathrm{ELC}$ extracted from Figs. \ref{figure5}a (and similar data for other samples) and \ref{figure7},
respectively, along with the phase boundaries of commensurate antiferromagnetism \cite{Coneri10} and superconductivity
\cite{Liang05}. Panel (b) shows a schematic of a $p$--$\textbf{H}$ section of the phase diagram derived in a
field-theoretical treatment of competing SDW and SC order parameters \cite{Demler01}. The control parameter in the
field theory was heuristically identified with the doping level $p$. The field range covered by the data of Fig.
\ref{figure4} and the larger field range covered by the quantum oscillation experiments on \ybcosixfive\ samples are
indicated by red arrows. } \label{figure8}
\end{center}
\end{figure}

Figure \ref{figure8}a summarizes the overall layout of the $p$--$T$ phase diagram of underdoped \ybco. The SDW and ELC
crossover lines established by our neutron scattering results are shown along with the previously determined phase
boundaries for commensurate antiferromagnetism \cite{Coneri10} and superconductivity \cite{Liang05}. At $T=0$, the SDW
and SC phases coexist over some range of $p$. Our data are consistent with $\mu$SR evidence of coexistence between
magnetic order and superconductivity in this doping regime \cite{Coneri10,Miller06}, but $\mu$SR does not contain
information about the spatial character of the local magnetization. The neutron scattering data presented here
demonstrate that the magnetic order coexisting with superconductivity is incommensurate. At the present time, they are
insufficient to determine whether the coexistence is spatially uniform, as predicted by spiral models
\cite{Sushkov05,Sushkov09}, or spatially modulated in the form of stripes \cite{Kivelson03,Berg07} or spin vortex
lattices \cite{Fine07}. In any case, unavoidable oxygen defects in the \ybco\ crystal lattice are expected to induce
some degree of spatial inhomogeneity.

Both the SDW order parameter extracted from the intensity of the quasielastic peak as $T \rightarrow 0$ and the
crossover temperature $T_\mathrm{SDW}$ that characterizes its onset at $T \neq 0$ are continuously reduced with
increasing $p$. This indicates that the SDW phase boundary ends at a quantum critical point (QCP) at $p \sim 0.1$,
close to the chemical composition \ybcosixfive. Further evidence for quantum criticality is derived from the scaling
properties of the dynamical spin correlations of \ybcosixfourfive\ \cite{Hinkov10}. In line with this scenario, the
scattering function at $T =0$ exhibits a spin gap for $p > 0.1$ \cite{Fong00}, but a dilute concentration of
nonmagnetic impurities restores the SDW \cite{Suchaneck10}.

The low-$p$ end of the incommensurate SDW regime is complicated by a confluence of several factors including the
tetragonal-to-orthorhombic transition \cite{Jorgensen90}, the onset of appreciable effective interactions between
CuO$_2$ bilayers ($\sim J_\perp \xi^2$, where $J_\perp \sim 0.02$ meV is the bare inter-bilayer interaction
\cite{Tranquada89} and $\xi$ is the intra-layer correlation length, which increases with decreasing $p$), and
dopant-induced disorder, which may lead to the formation of local Cu$^{2+}$ moments on the CuO chains for samples with
low oxygen content \cite{Kadowaki88}. Nonetheless, it is interesting to note that the onset of incommensurability in
the spin correlations (Fig. \ref{figure3}) nearly coincides with the onset of superconductivity (Fig. \ref{figure8}).
The \ybco\ system is less dominated by disorder than strongly underdoped \lsco\ \cite{Chen09} and thus provides an
interesting complementary forum for future exploration of the interplay between spin and charge correlations close to
the Mott insulating state.

In order to relate our low-temperature neutron scattering data to the results of the quantum oscillation experiments,
we have schematically drawn the range of magnetic fields covered by our experiments on top of the zero-temperature
phase diagram for coupled SDW and $d$-wave SC parameters derived from field theory \cite{Demler01} (Fig.
\ref{figure8}b). Here, the control parameter in the field theory was heuristically identified with the hole
concentration $p$, without claims to theoretical rigor or quantitative accuracy. Whereas \ybcosixthreefive\ appears to
be near the SDW end of the coexistence regime, far away from the QCP, both our \ybcosixfourfive\ sample and the
\ybcosixfive\ samples used in the quantum oscillation experiments are close to the QCP. The limited magnetic fields
currently available for neutron scattering are incapable of promoting the electron system into the uniform SDW phase
expected at high $\textbf{H}$, but sufficient to substantially increase the SDW order parameter in the coexistence
regime. In models with uniform SDW-SC coexistence for $\textbf{H}=0$, this proceeds via a spatially periodic
enhancement of the SDW order in the vicinity of magnetic vortices in the mixed state of the $d$-wave SC. It
will be interesting to explore a possible relation between the spatially modulated SDW order and the series of
field-induced transitions recently observed in the magnetic vortex lattice of optimally doped \ybco\ \cite{White09}.
The much higher fields used for the quantum oscillation experiments are expected to bring the electron system of
samples with doping levels near the zero-field QCP either close to the high-field boundary of the SDW-SC coexistence
range or into the uniform SDW phase (Fig. \ref{figure8}b). An extrapolation of our low-field measurements of the SDW
order parameter of \ybcosixfourfive\ to $\textbf{H} = 50$ T yields $\sim 0.13 \mu_B$ per Cu ion \cite{Haug09},
comparable to the zero-field sublattice magnetization of stripe-ordered Nd- or Eu-substituted \lsco, where recent
photoemission experiments have uncovered evidence of a reconstructed Fermi surface \cite{Chang08b,Zabolotnyy09}. It is
thus likely that a SDW reconstruction of the multiband Fermi surface of \ybcosixfive\ contributes to the formation
of the small pockets observed in the quantum oscillation experiments. This conclusion is supported by recent
quantitative assessments of the quantum oscillation data \cite{Harrison09,Sebastian10}. It is interesting to note that the QCP indicated by our
neutron scattering study coincides with the insulator-metal transition inferred from transport experiments in high
magnetic fields \cite{Sun04,Sebastian10b}, and that the transport data indicate a critical divergence of the cyclotron
mass at this doping level \cite{Sebastian10b,Norman10}. A comprehensive explanation of the neutron-scattering and
quantum-oscillation data therefore requires a transition from ``SDW insulator'' to ``SDW metal'' at high fields \cite{Sachdev10},
for which there is thus far no direct experimental evidence.

Going back to the zero-field phase diagram of Fig. \ref{figure8}a, we note that the ELC onset temperature also
increases with decreasing $p$, such that the ELC and SDW crossover lines are nearly parallel. In view of the minute
size ($\sim 0.5\%$ for \ybcosixthreefive) and opposite doping trend of the orthorhombic distortion of the crystal
structure \cite{Jorgensen90,Ando02}, this observation supports our contention \cite{Hinkov08} that the anisotropy is
not a simple consequence of the crystal symmetry \cite{Singh10}, but arises spontaneously from many-body interactions
between spins. The ELC onset temperature is substantially lower than the onset of the ``charge pseudogap'' detected by
infrared spectroscopy \cite{Timusk99,Yu08,Dubroka10}, which at these doping levels is well above room temperature.
However, it coincides with the onset of signatures of phase-incoherent superconductivity in the infrared spectra of
\ybco\ samples from the same batch as ours \cite{Dubroka10}. The theoretical description of the subtle interplay
between superconducting and spin correlations for $T \sim T_{\rm ELC}$ is an interesting subject of further
investigation, as is the relationship between the ELC phenomenon reported here and the spontaneous onset of in-plane
anisotropies in $dc$ resistivity \cite{Ando02} and Nernst effect \cite{Daou10} measurements. While the temperature
evolution of the in-plane resistivity anisotropy \cite{Ando02} is in good qualitative agreement with the anisotropy of
the spin dynamics shown in Fig. \ref{figure7}, corresponding Nernst-effect data are not yet available for the doping
levels investigated here.

Finally, it will be important to follow the ELC crossover line to higher doping levels, where the SDW phase is no
longer present at $T=0$. For \ybcosixsix\ ($p \sim 0.12$), the spin excitation spectrum exhibits a gap in the
superconducting state at $T=0$ \cite{Fong00}. Weak low-energy spin excitations are present above $T_\mathrm{c}$, but
their temperature dependence has not yet been studied in detail. The intensity of higher-energy excitations with $E
\sim 30$ meV, however, shows a strong upturn upon cooling below $\sim 150$ K \cite{Hinkov07}, in qualitative agreement
with transport \cite{Ando02,Daou10} and infrared \cite{Dubroka10} data. Taken together, these results suggest that the
ELC phenomenon persists at least up to $p \sim 0.12$. It will be interesting to explore the relationship between this
transition and the ``spin pseudogap'' detected by nuclear magnetic resonance \cite{Alloul09}, as well as the magnetic
reflections with $\textbf{Q} =0$ detected by elastic neutron scattering \cite{Fauque06,Mook08} in this doping regime.

\section*{Acknowledgements}

We thank O. Sushkov, G. Khaliullin, C. Bernhard, Ch. Niedermayer, A. Dubroka, Y. Pashkevich, D. Inosov, and Y. Li for discussions, S. Lacher, B. Baum
and H. Wendel for their help in sample preparation, C. Busch and H. Bender for technical support, and C. Stefani and R.
Dinnebier for determining the lattice parameters using X-ray powder diffraction. We acknowledge financial support by
the DFG under Grant No. FOR538.

\end{document}